\documentclass[reprint, twocolumn, showkeys]{revtex4}

\usepackage{graphicx}
\usepackage{dcolumn}
\usepackage{bm}
\usepackage{color}
\usepackage{amsmath,amsthm}
\usepackage{amssymb,bm}
\usepackage{mathrsfs}
\usepackage{comment}
\usepackage{ulem}

\begin{document}

\title{Molecular Electron Transfer in Optical Cavities: From Excitonic to Vibronic Polaritons}
\author{Takumi Hidaka}
\affiliation{Department of Molecular Engineering, Graduate School of Engineering, Kyoto University, Kyoto 615-8510, Japan}
\author{Tomohiro Fukushima}
\affiliation{Department of Chemistry, Faculty of Science, Hokkaido University, 060-0810 Sapporo, Japan}
\author{Nguyen Thanh Phuc}
\email{nthanhphuc@moleng.kyoto-u.ac.jp}
\affiliation{Department of Molecular Engineering, Graduate School of Engineering, Kyoto University, Kyoto 615-8510, Japan}

\begin{abstract}
Strong coupling between molecular excitations and quantized electromagnetic fields in optical cavities provides a powerful means to control the physical and chemical properties of molecular systems.
Here, we study electron transfer (ET) dynamics in cavity-coupled molecules using the numerically exact hierarchical equations of motion (HEOM) method, which captures nonperturbative and non-Markovian effects beyond standard perturbative theories.  
We identify distinct resonance and collective effects associated with polariton formation and show that the ET rate saturates in the strong-coupling regime, a feature not captured by perturbative approaches.  
We further extend the cavity-modified ET model by incorporating the nuclear-coordinate dependence of molecular electric dipole moments, which gives rise to a three-body interaction involving molecular electronic and vibrational degrees of freedom and cavity photons.
This vibronic polariton formation leads to non-monotonic, oscillatory dependencies of the ET rate on the light-matter coupling strength and cavity frequency, which we attribute to quantum interference among multiple transfer pathways.
These findings establish cavity-modified electron transfer as a multichannel quantum process governed by the interplay of electronic, vibrational, and photonic degrees of freedom.
\end{abstract}


\maketitle

\section{Introduction}
Recent advances in cavity quantum electrodynamics have established that confining molecules within tailored electromagnetic environments can strongly couple their internal excitations to quantized light fields, opening new routes to control material properties and chemical reactivity~\cite{Vidal21}.
In this strong-coupling regime, the hybridization of optical cavity modes with molecular excitations gives rise to light-matter eigenstates known as molecular polaritons~\cite{Ebbesen16, Nagarajan21}.
Depending on whether the cavity couples predominantly to electronic or vibrational molecular transitions, one distinguishes exciton polaritons and vibrational polaritons, respectively.
The latter, formed under vibrational strong coupling, has been reported to influence ground-state chemical reactivity~\cite{Thomas16, Lather19, Thomas19, Ahn23} and has been linked to enhanced intermolecular vibrational energy transfer~\cite{Xiang20} and accelerated intramolecular vibrational relaxation~\cite{Chen22}. 

Electronic strong coupling and the resulting exciton polaritons have been widely explored as platforms to reshape exciton migration, nonadiabatic dynamics, photochemical transformations~\cite{Bhuyan23, Pandya21, Pandya22, Cheng23, Xu23, Balasubrahmaniyam23, Zeng23, Peng23, Lee24, Amin24}, and spin transport~\cite{Phuc23}. 
Among the processes amenable to cavity control, condensed-phase electron transfer (ET) is of particular importance, as it underpins charge separation and transport in organic electronics, redox chemistry, and numerous biological functions~\cite{MayKuhnTextbook, NitzanTextbook}. 

In condensed-phase ET, the transfer rate is governed by the interplay among three key elements: (i) the electronic coupling between donor and acceptor states; (ii) the energetic driving force and activation barrier, shaped by solvent and intramolecular reorganization; and (iii) the thermal fluctuations and dissipation induced by the molecular environment. 
Optical cavities can, in principle, modify each of these elements. 
Hybridizing electronic states with a cavity mode creates polaritonic manifolds that reshape energy gaps and effective couplings. 
Additional radiative and dissipative channels introduced by cavity leakage alter the competition between coherent and incoherent dynamics. 
Furthermore, collective light-matter coupling in molecular ensembles can renormalize effective interactions and change how disorder and fluctuations impact reaction kinetics. 
Together, these mechanisms make cavity-modified ET a rich arena for controlling charge transport and chemical reactivity in nanophotonic environments~\cite{Hayashi24}. 
Despite growing theoretical interest in cavity-modified condensed-phase ET~\cite{Semenov19, Phuc20, Mandal20, Chowdhury22, Wei22, Saller22, Saller23, SallerJCP23, Koessler25}, most existing studies have relied on perturbative rate expressions, typically within the Fermi's golden rule (FGR) framework, which are rigorously justified only in the linear response regime. 
However, experimentally relevant cavity QED settings can involve strong light-matter coupling and structured vibrational environments, where nonperturbative and non-Markovian effects may be essential for accurate quantitative predictions and mechanistic understanding.

In this work, we employ the hierarchical equations of motion (HEOM)~\cite{Tanimura20} to simulate the quantum dynamics of cavity-modified ET beyond the perturbative regime.
Effective transfer rates are extracted by exponential fitting of the time-evolving donor and acceptor populations computed from HEOM.
We organize our investigation as follows.
We begin with the minimal model of cavity-modified ET, in which the nuclear-coordinate dependence of the molecular electric dipole moments is neglected, so that the cavity couples exclusively to the electronic degrees of freedom. 
Within this model, we characterize ET rate modification as a function of the driving force and reorganization energy for two distinct coupling channels: direct transition coupling and energy fluctuation coupling. 
We examine the effect of strong light-matter coupling on coherence dynamics by comparing the system's time evolution with and without cavity coupling.
Resonance and collective effects arising from polariton formation are investigated by varying the cavity frequency and the number of molecules coupled to a single cavity mode, respectively. 
The role of cavity loss is assessed by augmenting the HEOM with a Lindblad-form dissipation term.
We then generalize the model to include the nuclear-coordinate-dependent molecular dipole moments, which introduces a three-body interaction coupling the molecular electronic and vibrational degrees of freedom to the cavity photons.
The influence of the vibronic polariton formation on ET dynamics is investigated for environments characterized by both Drude-Lorentz and underdamped spectral densities, revealing a rich interplay of quantum interference and nonlinear coupling effects.

\section{System Model}
We consider $N_\text{mol}$ identical molecules in the condensed phase strongly coupled to a single mode of an optical cavity. 
In the dipole gauge, the light-matter interaction is described by the Hamiltonian~\cite{Semenov19}:
\begin{align}
    \hat{H}=&\sum_{n=1}^{N_\text{mol}}\left(\hat{H}_\text{M}^{(n)}+\hat{H}_\text{B}^{(n)}+\hat{H}_\text{MB}^{(n)}\right)
    +\hbar\omega_\text{c}\hat{a}^\dagger\hat{a}\nonumber\\
    &-\sum_{n=1}^{N_\text{mol}}\hat{\boldsymbol{\mu}}^{(n)}\cdot\hat{\mathbf{E}} 
    +\frac{1}{2\mathcal{V}\epsilon_0} \left(\sum_{n=1}^{N_\text{mol}}\hat{\boldsymbol{\mu}}^{(n)}\cdot\boldsymbol{\epsilon}\right)^2,
    \label{eq: Hamiltonian in dipole gauge}
\end{align}
where $\hat{H}_\text{M}^{(n)}$, $\hat{H}_\text{B}^{(n)}$, and $\hat{H}_\text{MB}^{(n)}$ represent the molecular, bath (environment), and molecule-bath interaction Hamiltonians, respectively, of the $n$-th molecule.
The cavity mode, characterized by frequency $\omega_\text{c}$, polarization unit vector $\boldsymbol{\epsilon}$, and effective mode volume $\mathcal{V}$, is described by photonic creation and annihilation operators $\hat{a}^\dagger$ and $\hat{a}$. 
Light-matter interaction enters through the coupling between the total molecular electric dipole moment $\hat{\boldsymbol{\mu}}^{(n)}$ and the cavity electric field operator
\begin{align}
    \hat{\mathbf{E}}=i\boldsymbol{\epsilon}\sqrt{\frac{\hbar\omega_\text{c}}{2\mathcal{V}\epsilon_0}}
    \left(\hat{a}-\hat{a}^\dagger\right),
\end{align} 
where $\epsilon_0$ is the vacuum permittivity. 
The final term in Eq.~\eqref{eq: Hamiltonian in dipole gauge} is the dipole self-energy, which ensures gauge invariance and becomes significant in the ultrastrong coupling regime where the light-matter coupling strength is comparable to the characteristic molecular energy scales and the cavity frequency~\cite{FriskKockum19}. 
This dipole gauge Hamiltonian is obtained via the Power-Zienau-Woolley unitary transformation from the minimal coupling Hamiltonian in the Coulomb gauge~\cite{TannoudjiTextbook}.

We first consider a single molecule ($N_\text{mol}=1$) strongly coupled to a single cavity mode. 
To model electron transfer, we project the molecular Hamiltonian onto the two relevant electronic states: the donor state $|\text{D}\rangle$ and acceptor state $|\text{A}\rangle$. 
The environment is represented as a collection of harmonic oscillators with frequencies $\nu_j$ and annihilation operators $\hat{b}_j$, yielding the combined molecular and bath Hamiltonian:
\begin{align}
    &\hat{H}_\text{M}+\hat{H}_\text{B}+\hat{H}_\text{MB}\nonumber\\
    =&E_\text{D}|\text{D}\rangle\langle\text{D}|
    +\left[E_\text{A}+\sum_j \lambda_j(\hat{b}_j+\hat{b}_j^\dagger)\right]|\text{A}\rangle\langle\text{A}|\nonumber\\
    &+\sum_j\hbar\nu_j\hat{b}_j^\dagger\hat{b}_j 
    +H_\text{DA}|\text{D}\rangle\langle\text{A}|+H_\text{AD}|\text{A}\rangle\langle\text{D}|,
\end{align}
where $E_\text{D}$ and $E_\text{A}$ are the donor and acceptor energies, $H_\text{DA}$ and $H_\text{AD}$ characterize the electronic coupling between states, and $\lambda_j$ represents the vibronic coupling strength of the $j$-th bath mode.
Throughout our numerical calculations, we set $H_\text{DA}=H_\text{AD}=30\,\text{cm}^{-1}$.

Expressing the full Hamiltonian in the donor-acceptor basis in terms of dipole matrix elements yields~\cite{Semenov19}
\begin{align}
    \hat{H}=&E_\text{D}|\text{D}\rangle\langle\text{D}|
    +\left[E_\text{A}+\sum_j \lambda_j(\hat{b}_j+\hat{b}_j^\dagger)\right]|\text{A}\rangle\langle\text{A}|\nonumber\\
    &+\sum_j\hbar\nu_j\hat{b}_j^\dagger\hat{b}_j 
    +H_\text{DA}|\text{D}\rangle\langle\text{A}|+H_\text{AD}|\text{A}\rangle\langle\text{D}|
    +\hbar\omega_\text{c}\hat{a}^\dagger\hat{a}\nonumber\\
    &+\hbar\omega_\text{c} \left(\hat{a}-\hat{a}^\dagger\right)\Big(g_\text{D}|\text{D}\rangle\langle\text{D}|+ g_\text{A}|\text{A}\rangle\langle\text{A}|\nonumber\\
    &+t_\text{DA}|\text{D}\rangle\langle\text{A}|+t_\text{AD}|\text{A}\rangle\langle\text{D}|\Big)
    +\hbar\omega_\text{c}\Big[|g_\text{D}|^2|\text{D}\rangle\langle\text{D}|\nonumber\\
    &+|g_\text{A}|^2|\text{A}\rangle\langle\text{A}|
    +|t_{DA}|^2
    -\left(g_\text{D}+g_\text{A}\right)\Big(t_\text{DA}|\text{D}\rangle\langle\text{A}|\nonumber\\
    &+t_\text{AD}|\text{A}\rangle\langle\text{D}|\Big)\Big],
\label{eq: Hamiltonian in the donor-acceptor basis}
\end{align}
where the coupling parameters are defined by projections of the molecular dipole operator onto the donor-acceptor basis:
\begin{align}
    g_\text{D}=&i\sqrt{\frac{1}{2\hbar\omega_\text{c}\mathcal{V}\epsilon_0}}\boldsymbol{\mu}_\text{DD}\cdot\boldsymbol{\epsilon},\label{eq: g_D}\\
    g_\text{A}=&i\sqrt{\frac{1}{2\hbar\omega_\text{c}\mathcal{V}\epsilon_0}}\boldsymbol{\mu}_\text{AA}\cdot\boldsymbol{\epsilon},\\
    t_\text{DA}=&i\sqrt{\frac{1}{2\hbar\omega_\text{c}\mathcal{V}\epsilon_0}}\boldsymbol{\mu}_\text{DA}\cdot\boldsymbol{\epsilon} \label{eq: t_DA}
\end{align}
with $\boldsymbol{\mu}_\text{DD}=\langle\text{D}|\hat{\boldsymbol{\mu}}|\text{D}\rangle$, $\boldsymbol{\mu}_\text{AA}=\langle\text{A}|\hat{\boldsymbol{\mu}}|\text{A}\rangle$, and $\boldsymbol{\mu}_\text{DA}=\langle\text{D}|\hat{\boldsymbol{\mu}}|\text{A}\rangle$ denoting the diagonal (permanent) and off-diagonal (transition) dipole matrix elements.
In the minimal model formulation, we treat the dipole matrix elements as constants, independent of nuclear coordinates. 
This approximation is valid when the Born-Oppenheimer separation holds and the electronic charge distribution is only weakly modulated by vibrational motion.
We will relax this constraint in the generalized model presented later, where nuclear-coordinate dependence of the dipole moments gives rise to a three-body electronic-vibrational-photonic interaction. 
The final term in Eq.~\eqref{eq: Hamiltonian in the donor-acceptor basis} accounts for dipole self-energy contributions.

This formulation reveals two distinct coupling channels through which the cavity mode influences electron transfer. 
The first channel, characterized by $g_\text{D}$ and $g_\text{A}$, corresponds to cavity-induced fluctuations of the donor and acceptor state energies (energy-fluctuation coupling).
The second channel, characterized by $t_\text{DA}=-t_\text{AD}^*$, couples the cavity field directly to the electronic transition between donor and acceptor states (direct transition coupling).  

The influence of the environment on the system dynamics is encoded in the spectral density $J(\omega)=(2\pi/\hbar)\sum_j \lambda_j^2\delta(\omega-\nu_j)$, with the overall system-environment coupling strength quantified by the reorganization energy $\lambda_0=\sum_j \lambda_j^2/(\hbar\nu_j)$.
Unless otherwise stated, we employ a Drude-Lorentz spectral density representing a collection of low-frequency intermolecular vibrational modes:
\begin{align}
	J_\text{DL}(\omega)=\frac{4\lambda_0 W\omega}{\omega^2+W^2},
\end{align} 
where the cutoff frequency $W=\tau_\text{b}^{-1}$ is inversely related to the bath relaxation time, which we set to $\tau_\text{b}=100\;\text{fs}$.

To capture the system's quantum dynamics, we employ the hierarchical equations of motion (HEOM)~\cite{Tanimura20}, which provides numerically exact solutions for open quantum systems across arbitrary coupling strengths. 
The HEOM approach proceeds by first expanding the bath correlation function
\begin{align}
    C(t)=\int_0^\infty \text{d}\omega J(\omega)\left[\coth\left(\frac{\beta\omega}{2}\right)\cos\omega t-i\sin\omega t\right],
\end{align}
into a sum of exponential functions, exploiting the structure of the spectral density $J(\omega)$ and temperature-dependent factors. 
After formally integrating out the bath degrees of freedom, one obtains a hierarchical set of coupled differential equations governing not only the reduced density matrix of the system but also a hierarchy of auxiliary density operators (ADOs). 
These ADOs encode the complete memory of past system-bath interactions at different time depths. 
Each level of the hierarchy captures correlations extending further back in time, with higher-level ADOs representing increasingly long-lived memory effects. 
The coupled evolution of the system density operator and the ADO hierarchy enables exact treatment of non-Markovian dynamics, capturing memory effects that persist when the bath relaxation timescale is comparable to or longer than the characteristic timescales of the system's evolution.

We solve the HEOM to obtain the time evolution of the donor and acceptor populations, $p_\text{D}(t)$ and $p_\text{A}(t)=1-p_\text{D}(t)$, with initial conditions corresponding to the molecule in the donor state and the environment in thermal equilibrium: $p_\text{D}(0)=1$ and $p_\text{A}(0)=0$. 
Effective transfer rates are extracted by fitting the computed population dynamics to the kinetic rate equations:
\begin{align}
    \frac{\text{d}p_\text{D}}{\text{d}t}=&-k_\text{DA}p_\text{D}+k_\text{AD}p_\text{A}, \label{eq: equation 1}\\
    \frac{\text{d}p_\text{A}}{\text{d}t}=&k_\text{DA}p_\text{D}-k_\text{AD}p_\text{A}, \label{eq: equation 2}
\end{align}
where $k_\text{DA}$ and $k_\text{AD}$ denote the forward and backward transfer rates. 
The solution to these equations yields
\begin{align}
    p_\text{D}(t)=\frac{k_\text{AD}}{k_\text{AD}+k_\text{DA}}
    +\frac{k_\text{DA}}{k_\text{AD}+k_\text{DA}}e^{-(k_\text{AD}+k_\text{DA})t},
\end{align}
from which $k_\text{DA}$ is determined via exponential fitting.

\section{Results and Discussion}
\subsection{Minimal Model: Cavity-Electronic Coupling}
\begin{figure*}
	\centering
	\includegraphics[width=0.9\textwidth]{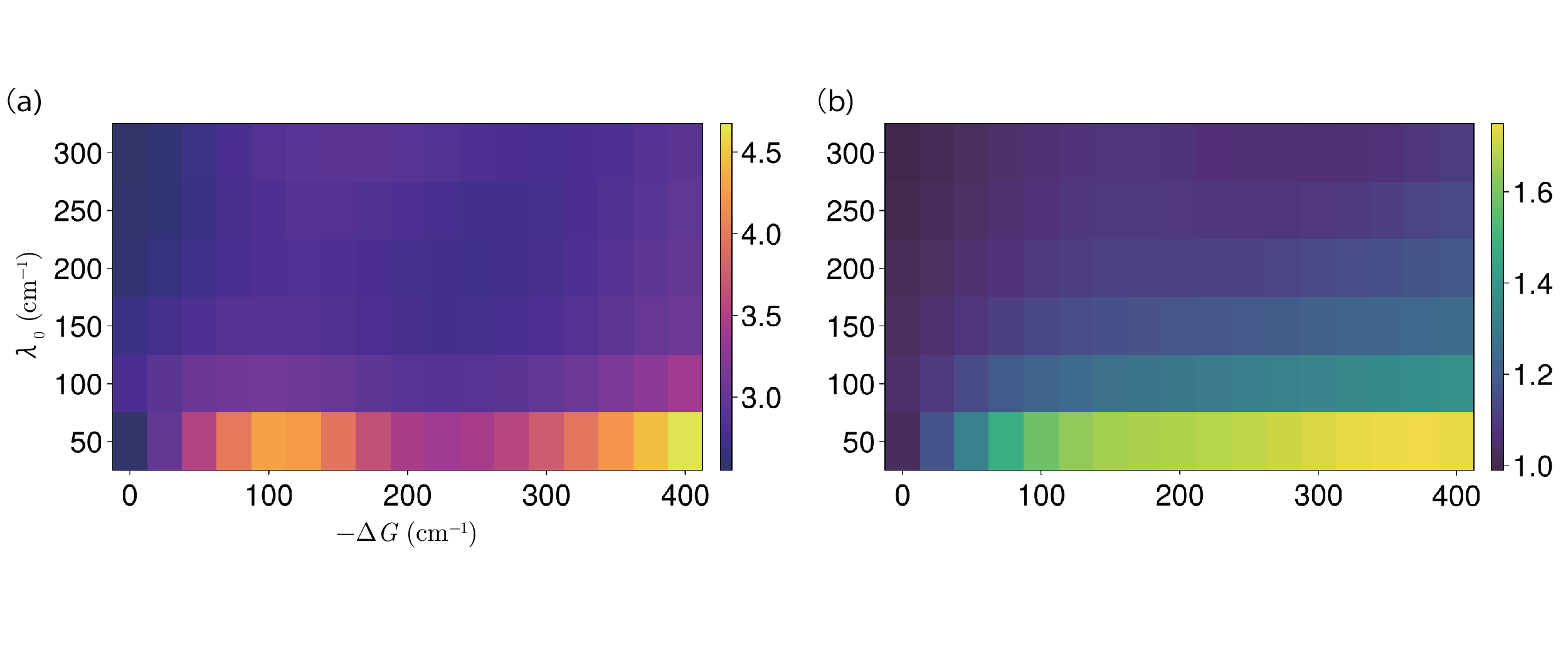}
	\caption{Relative change $k_\text{cav}/k_\text{non-cav}$ in electron transfer rate as a function of reorganization energy $\lambda_0$ and driving force $-\Delta G$ for a single molecule ($N_\text{mol}=1$). (a) Direct transition coupling only ($t_\text{DA}\neq 0$, $g_\text{D}=g_\text{A}=0$). (b) Energy-fluctuation coupling only ($t_\text{DA}=0$, $g_\text{D,A}\neq 0$).}
	\label{fig: Heatmap Plots}
\end{figure*}

Figure~\ref{fig: Heatmap Plots} displays the relative change in electron transfer rate, $k_\text{cav}/k_\text{non-cav}$, as a function of the reorganization energy $\lambda_0$ and the driving force $-\Delta G=E_\text{D}-E_\text{A}$.
Panel (a) shows results when light-matter interaction occurs exclusively through the direct transition coupling channel ($t_\text{DA}\neq 0$, $g_\text{D}=g_\text{A}=0$). 
Here we set $\hbar\omega_\text{c}=100\,\text{cm}^{-1}$ and $|t_\text{DA}|=0.5$. 
The cavity coupling produces substantial rate enhancement across a broad parameter space, with increases exceeding four-fold under optimal conditions. 
This pronounced effect arises because the cavity field couples directly to the transition dipole moment, effectively opening a new coherent pathway for electron transfer.
Panel (b) presents the corresponding results when interaction occurs solely through the energy-fluctuation coupling channel ($t_\text{DA}=0$, $g_\text{D,A}\neq 0$), with parameters $\hbar\omega_\text{c}=100\,\text{cm}^{-1}$, $|g_\text{D}|=0.6$, and $|g_\text{A}|=0.1$.
While this channel also yields rate enhancement, the effect is less pronounced than for the $t_\text{DA}$ channel. 
Here, the cavity modulates the energies of the donor and acceptor states rather than directly facilitating the electronic transition, resulting in a qualitatively different enhancement mechanism with generally smaller magnitude.

\begin{figure}
	\centering
	\includegraphics[width=0.9\linewidth]{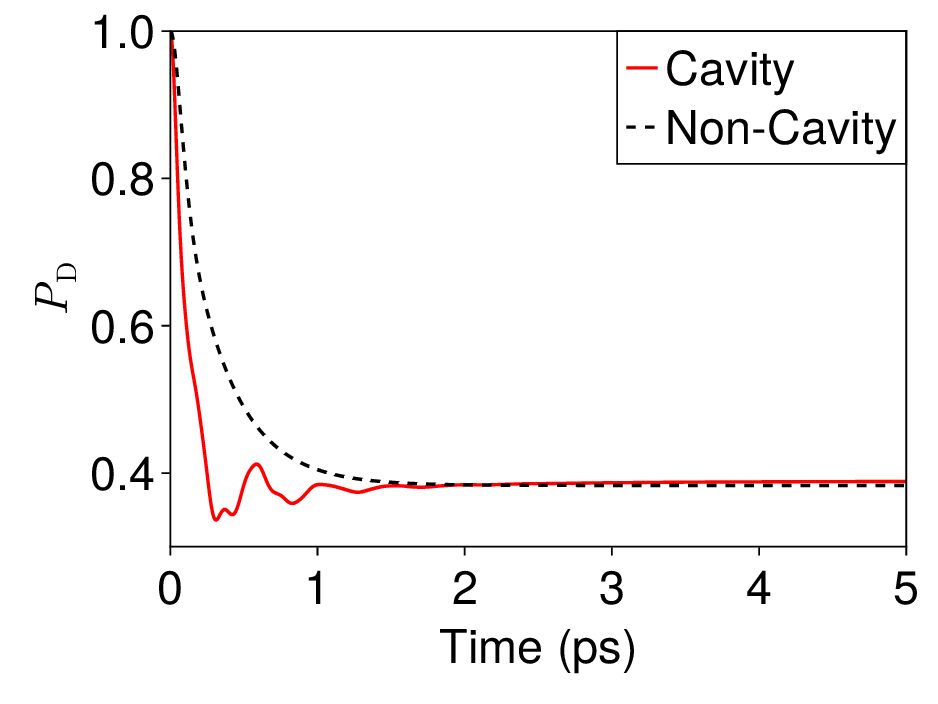}
	\caption{Time evolution of the donor population $p_\text{D}(t)$ for the direct transition coupling with $\lambda_0=50\,\text{cm}^{-1}$ and $-\Delta G=100\,\text{cm}^{-1}$. Red solid line: cavity-coupled system ($\hbar\omega_\text{c}|t_\text{DA}|=50\,\text{cm}^{-1}$). Black dashed line: uncoupled system. Cavity coupling accelerates transfer and extends coherence time through the dynamic polaron decoupling effect.}
	\label{fig: population dynamics-tchannel}
\end{figure}

To examine how cavity coupling affects the temporal dynamics of electron transfer, we plot the time evolution of the donor population in Fig.~\ref{fig: population dynamics-tchannel} for the direct transition coupling channel with $\lambda_0=50\,\text{cm}^{-1}$ and $-\Delta G=100\,\text{cm}^{-1}$. 
Comparison between cavity-coupled (red solid) and uncoupled (black dashed) cases reveals two distinct effects of the light-matter interaction. 
First, the cavity accelerates the transfer process, as evidenced by the faster decay of the donor population toward its equilibrium value. 
Second, and equally important, the cavity-coupled system exhibits pronounced oscillatory behavior that persists over significantly longer timescales, indicating an extended coherence time compared to the bare molecular system. 

This prolonged coherence can be understood through the dynamic polaron decoupling effect~\cite{Phuc21, Phuc25, Spano15, Herrera16, Phuc19a, Takahashi20}. 
When light-matter coupling is sufficiently strong, the formation of delocalized polariton states effectively reduces the local system-environment interaction. 
The polariton, being a hybrid light-matter excitation distributed across both photonic and molecular components, is less susceptible to environmental fluctuations that would otherwise rapidly dephase purely molecular transitions. 
This decoherence protection mechanism allows coherent dynamics to survive longer, manifesting as sustained oscillations in the population dynamics.  

To assess the validity of perturbative approaches, we compare our numerically exact HEOM results with analytic expressions derived from Fermi's golden rule (FGR)~\cite{Semenov19}. 
FGR provides a widely used perturbative framework for calculating transition rates, treating the donor-acceptor coupling to second oder. 
For cavity-modified ET, the FGR expressions take different forms depending on the coupling channels. 
For the energy-fluctuation coupling channel ($g_\text{D,A}$ channel), the FGR rate is
\begin{align}
	k_\text{DA}^\text{FGR}=&\sqrt{\frac{\pi}{\hbar^2 k_\text{B}T\lambda_0}}
	H_\text{DA}^2 e^{-|g_\text{D}-g_\text{A}|^2}
	\sum_{m=0}^{\infty} \frac{|g_\text{D}-g_\text{A}|^{2m}}{m!}\nonumber\\
	&\times\exp\left\{-\frac{(-\Delta G-\lambda_0-m\hbar\omega)^2}{4k_\text{B}T\lambda_0}\right\}, 
	\label{eq: analytic expression,g-channel}
\end{align}
where the summation over photon number states $m$ reflects the cavity-induced modulation of the activation barrier. 
For the direct transition coupling channel ($t_\text{DA}$ channel), the FGR rate is
\begin{align}
	k_\text{DA}^\text{FGR}=&\sqrt{\frac{\pi}{\hbar^2 k_\text{B}T\lambda_0}}
	\Bigg[H_\text{DA}^2\exp\left\{-\frac{(-\Delta G-\lambda_0)^2}{4k_\text{B}T\lambda_0}\right\} \nonumber\\
	&+|\hbar\omega t_\text{DA}|^2\exp\left\{-\frac{(-\Delta G-\lambda_0-\hbar\omega)^2}{4k_\text{B}T\lambda_0}\right\}\Bigg],
	\label{eq: analytic expression, t-channel}
\end{align}
where the two terms represent the bare molecular transfer and the cavity-mediated transfer pathway, respectively. 

\begin{figure*}
	\centering
	\includegraphics[width=0.9\textwidth]{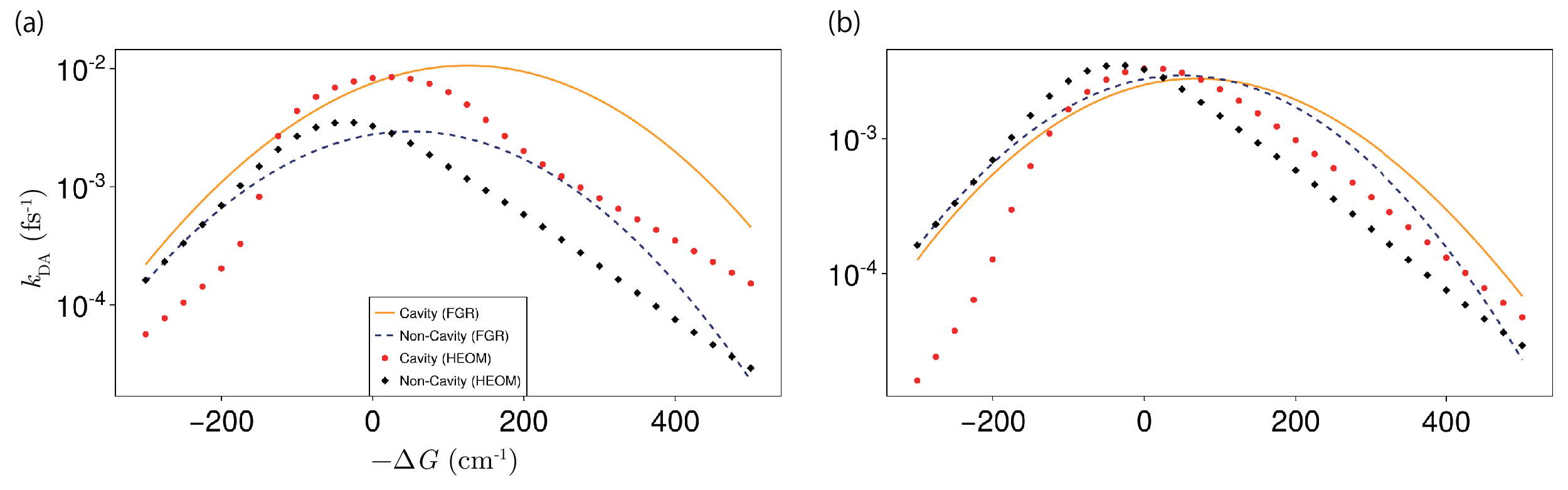}
	\caption{Comparison of ET rates as a function of driving force $-\Delta G$ for (a) the direct transition coupling channel ($t_\text{DA}$) and (b) the energy-fluctuation coupling channel ($g_\text{D,A}$), with $\lambda_0=50\,\text{cm}^{-1}$. Solid lines: Fermi's golden rule predictions. Dots: numerically exact HEOM results. While FGR captures qualitative trends, it fails to reproduce the quantitative rates and spectral widths due to neglect of nonperturbative and non-Markovian effects.}
	\label{fig: DeltaG-dependence: HEOM vs FGR}
\end{figure*}

Figure~\ref{fig: DeltaG-dependence: HEOM vs FGR} compares these FGR predictions (solid lines) with HEOM results (dots) as a function of driving force $-\Delta G$ for $\lambda_0=50\,\text{cm}^{-1}$. 
Although FGR reproduces the overall dependence of the ET rate on the driving force, the HEOM results exhibit systematically narrower profiles and a noticeable shift of the peak position relative to the FGR curves.
These discrepances stem from the perturbative nature of FGR. 
In contrast, HEOM fully captures nonperturbative couplings as well as non-Markovian memory effects that arise when the bath relaxation timescale is comparable to the characteristic timescale of electron transfer dynamics. 
These memory effects manifest as correlations between the system's past and present states, leading to rate modifications and spectral narrowing that cannot be predicted by time-local, perturbative theories.

\begin{figure}
	\centering
	\includegraphics[width=0.9\linewidth]{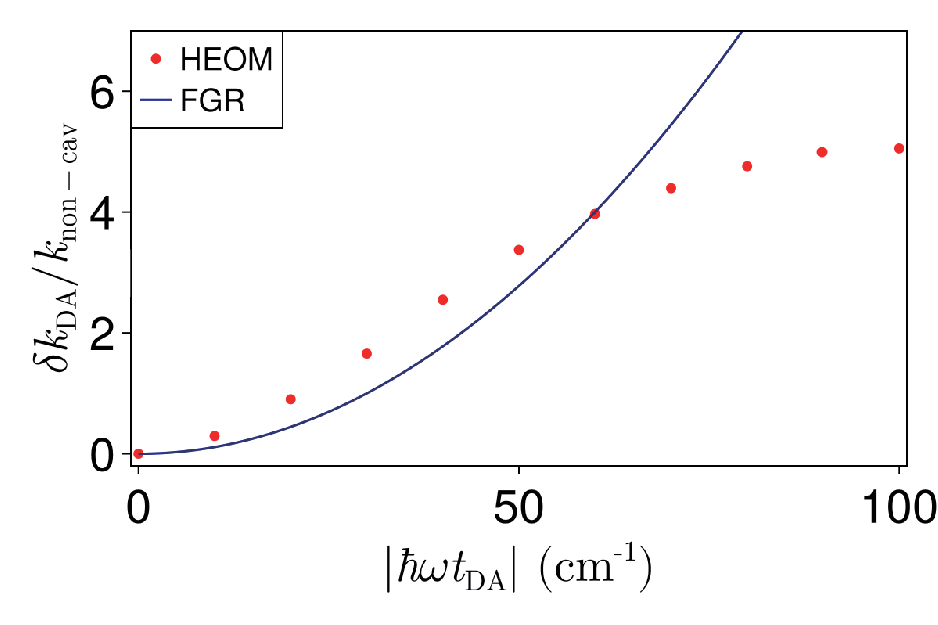}
	\caption{Cavity-induced change in ET rate, $\delta k_\text{DA}=k_\text{cav}-k_\text{non-cav}$ (normalized by $k_\text{non-cav}$) as a function of direct transition coupling strength $|t_\text{DA}|$ for $\lambda_0=50\,\text{cm}^{-1}$ and $-\Delta G=100\,\text{cm}^{-1}$. Red dots: HEOM results. Blue solid curve: FGR prediction (Eq.~\eqref{eq: analytic expression, t-channel}). The quadratic scaling $\delta k_\text{DA}\propto |t_\text{DA}|^2$ breaks down beyond $|t_\text{DA}|\sim 0.3$, where the rate saturates as the system enters the ultrastrong coupling regime.}
	\label{fig: tDA dependence}
\end{figure}

The limitations of perturbative theory become most pronounced when examining the dependence of the ET rate on light-matter coupling strength. 
According to Eq.~\eqref{eq: analytic expression, t-channel}, FGR predicts a quadratic scaling of the cavity-induced rate enhancement: $\delta k_\text{DA}\propto |t_\text{DA}|^2$, where $\delta k_\text{DA}=k_\text{cav}-k_\text{non-cav}$ represents the change in transfer rate upon cavity coupling. 
Figure~\ref{fig: tDA dependence} compares this FGR prediction (blue curve) with HEOM results (red dots) as a function of the coupling strength $|t_\text{DA}|$. 
The quadratic scaling holds reasonably well only in the coupling regime $|t_\text{DA}|\lesssim 0.3$ (corresponding to $\hbar\omega_\text{c}|t_\text{DA}|\lesssim 30\,\text{cm}^{-1}$). 
Beyond this threshold, the HEOM results deviate systematically from the perturbative prediction: the rate enhancement saturates rather than continuing to grow quadratically. 
This saturation signals entry into the ultrastrong coupling regime, where the light-matter interaction energy becomes comparable to other characteristic energy scales in the system.
Notably, even within the moderate coupling window where quadratic scaling approximately holds, the HEOM-predicted coefficient differs quantitatively from the FGR value. 
This discrepancy reflects the fact that FGR neglects non-Markovian bath correlations and higher-order virtual processes that contribute to the effective rate even at modest coupling strengths. 

\subsection{Resonance, Loss, and Collective Effects}
\begin{figure}
	\centering
	\includegraphics[width=0.9\linewidth]{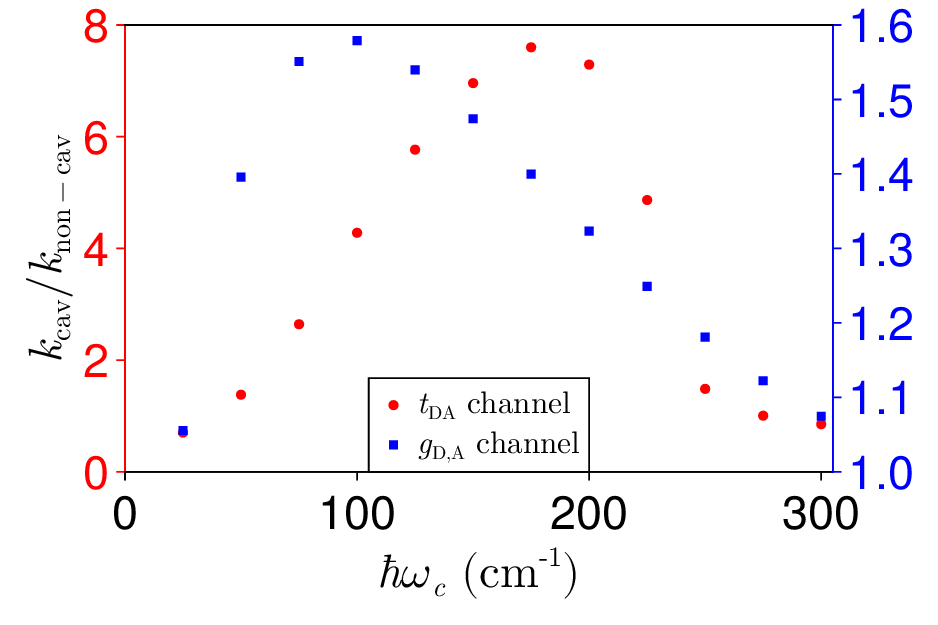}
	\caption{Resonance behavior of the ET rate as a function of cavity frequency $\omega_\text{c}$ with coupling parameters rescaled as $t_\text{DA}, g_\text{D,A}\propto 1/\sqrt{\omega_\text{c}}$ at fixed mode volume. (a) Direct transition coupling channel ($t_\text{DA}$): cavity-transition resonance. (b) Energy-fluctuation coupling channel ($g_\text{D,A}$): resonance reflects optimal cavity-induced modulation of the activation barrier. The different resonance frequencies underscore the distinct physical mechanisms of the two coupling channels.}
	\label{fig: dependence on omega_c}
\end{figure}

An important feature of cavity-modified ET is the resonance behavior that emerges when the cavity frequency is tuned relative to the molecular and environmental energies. 
Figure~\ref{fig: dependence on omega_c} shows the ET rate as a function of cavity frequency $\omega_\text{c}$ for both coupling channels. 
To isolate the effect of cavity frequency from changes in coupling strength, we rescale the coupling parameters according to $t_\text{DA}, g_\text{D,A}\propto 1/\sqrt{\omega_\text{c}}$ (see Eqs.~\eqref{eq: g_D}--\eqref{eq: t_DA}), keeping the effective light-matter interaction fixed at constant cavity mode volume $\mathcal{V}$. 
This normalization ensures that variations in the ET rate arise solely from energetic resonance conditions rather than from trivial changes in coupling magnitude. 
Both coupling channels exhibit clear resonance peaks where the ET rate is maximized, but the optimal frequencies differ significantly between the two mechanisms. 
For the direct transition coupling channel [panel (a)], the resonance occurs when the cavity frequency matches the effective energy gap between the donor and acceptor states.
At this condition, the cavity mode is resonant with the electronic transition, enabling efficient cavity-mediated transfer through coherent emission and reabsorption of virtual photons. 
In contrast, for the energy-fluctuation coupling channel [panel (b)], the resonance condition is more subtle, as it depends on the energetic overlap between cavity-shifted donor and acceptor potentials modulated by the reorganization energy and thermal fluctuations. 
The cavity effectively renormalizes the energetics of both states, and the optimal frequency reflects a balance between activation barrier reduction and thermal accessibility. 
These distinct resonance behaviors highlight the qualitatively different mechanisms by which the two coupling channels modify electron transfer: the $t_\text{DA}$ channel opens a new coherent transfer pathway through direct cavity-mediated coupling, while the $g_\text{D,A}$ channel reshapes the free energy landscape governing the transfer process.

\begin{figure}
	\centering
	\includegraphics[width=0.9\linewidth]{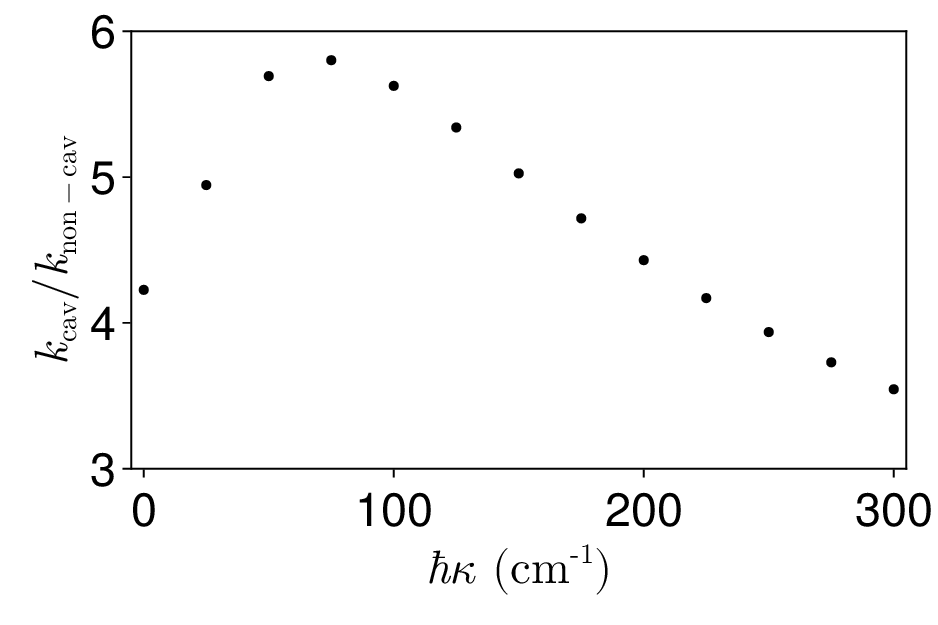}
	\caption{ET rate enhancement $k_\text{cav}/k_\text{non-cav}$ as a function of cavity loss rate $\kappa$ for the direct transition coupling channel ($t_\text{DA}$) with $\lambda_0=50\,\text{cm}^{-1}$ and $-\Delta G=100\,\text{cm}^{-1}$. The non-monotonic behavior reflects a competition between fluctuation-assisted transport in the low-dissipation regime and coherence suppression in the high-dissipation regime. The optimal enhancement occurs at intermediate loss rates where cavity-mediated coherent dynamics and dissipative broadening are balanced.}
	\label{fig: cavity loss}
\end{figure}

In realistic experimental implementations, optical cavities have finite quality factors due to photon leakage into external electromagnetic modes.
This cavity loss, characterized by a photon lifetime $\tau_\text{c}$, causes spectral broadening of the cavity resonance and introduces dissipation into the light-matter dynamics.
To capture these effects, we extend our model by incorporating Lindblad dissipation into the quantum master equation for the reduced density operator $\hat{\rho}$:
\begin{align}
\frac{\text{d}\hat{\rho}}{\text{dt}}=-\frac{i}{\hbar}\left[\hat{H},\hat{\rho}\right]
+\frac{\kappa}{2}\left(2\hat{a}\hat{\rho}\hat{a}^\dagger-\hat{a}^\dagger\hat{a}\hat{\rho}-\hat{\rho}\hat{a}^\dagger\hat{a}\right),
\end{align}
where $\kappa=\tau_\text{c}^{-1}$ is the cavity loss rate and the Lindblad term describes Markovian photon leakage from the cavity mode.
This dissipator models the irreversible decay of cavity photons while preserving the positivity and trace of the density operator, as required for a physically valid quantum evolution. 
Figure~\ref{fig: cavity loss} reveals a non-monotonic dependence of the ET rate enhancement on the cavity loss rate $\kappa$.
This behavior reflects a competition between two opposing effects. 
In the low-dissipation regime, increasing cavity loss actually enhances the ET rate. 
Here, dissipation plays a constructive role by broadening the cavity resonance, which relaxes stringent energy-matching requirements and enables fluctuation-assisted transport. 
This mechanism is analogous to environment-assisted quantum transport observed in other contexts, where moderate decoherence can facilitate incoherent hopping between states that would otherwise remain energetically misaligned~\cite{Rebentrost09, Phuc18, Phuc19b, Phuc23Floquet}. 
The photon leakage effectively provides an additional pathway for energy dissipation that can help the system overcome activation barriers. 
However, in the high-dissipation regime, excessive cavity loss becomes detrimental. 
Strong damping suppresses the coherent light-matter dynamics necessary for polariton formation and cavity-mediated transfer. 
The rapid photon decay prevents the buildup of coherent cavity excitation, essentially reverting the system toward the bare molecular limit.
This inhibition resembles a quantum Zeno-like effect: frequent dissipative events disrupt the coherent evolution before it can facilitate electron transfer. 
The optimal cavity loss rate, corresponding to the peak in Fig.~\ref{fig: cavity loss}, represents a balance between these competing mechanisms: sufficient dissipation to enable flexible energy matching, but not so much as to destroy the coherent polariton dynamics underlying cavity-enhanced transfer.

A distinctive feature of cavity QED is the emergence of collective effects when multiple emitters couple to a shared cavity mode. 
Whether such collective coupling can influence electron transfer---a fundamentally local process occurring at individual molecular sites---is a non-trivial question, especially when the per-molecule light-matter coupling strength is modest.
The answer depends on whether collective polariton formation modifies the energy landscape and decoherence dynamics in ways that align favorably or unfavorably with the conditions for efficient ET.
To investigate collective effects systematically, we compare systems with $N_\text{mol}=1$ and $N_\text{mol}=2$ molecules while maintaining fixed per-molecule coupling strength.
This ensures that the light-matter interaction per molecule remains constant, allowing us to isolate purely collective phenomena arising from cooperative coupling to the same cavity mode. 
All molecules are initialized in their donor states with the cavity in the vacuum state. 
In the absence of cavity coupling, ET occurs independently at each molecule, and the rate is naturally independent of $N_\text{mol}$. 
We neglect the dipole self-energy term in this analysis to focus on the essential collective physics. 

\begin{figure}
	\centering
	\includegraphics[width=0.9\linewidth]{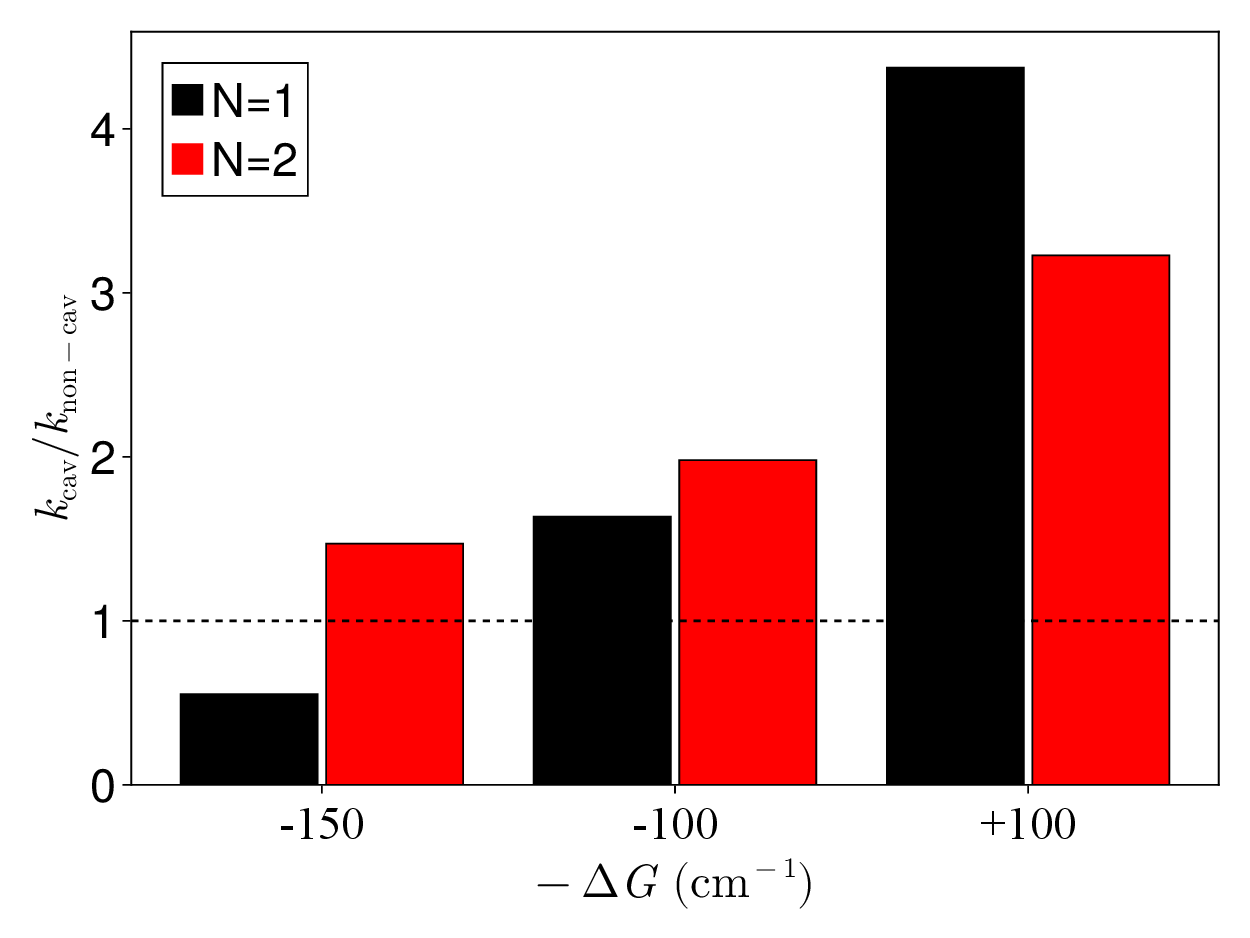}
	\caption{Parameter-dependent collective effects in cavity-modified ET for the direct transition coupling channel. Plotted is the rate enhancement $k_\text{cav}/k_\text{non-cav}$ comparing single-molecule ($N_\text{mol}=1$, black) and two-molecule ($N_\text{mol}=2$, red) systems at fixed per-molecule coupling strength for three driving forces: (center) $-\Delta G=-100\,\text{cm}^{-1}$ showing positive collective effect with $k_\text{non-cav}<k_\text{cav}(N_\text{mol}=1)<k_\text{cav}(N_\text{mol}=2)$; (right) $-\Delta G=100\,\text{cm}^{-1}$ showing negative collective effect with $k_\text{non-cav}<k_\text{cav}(N_\text{mol}=2)<k_\text{cav}(N_\text{mol}=1)$; (left) $-\Delta G=-150\,\text{cm}^{-1}$ showing transition behavior with $k_\text{cav}(N_\text{mol}=1)<k_\text{non-cav}<k_\text{cav}(N_\text{mol}=2)$. The sign and magnitude of collective effects depend on whether collective Rabi splitting and reduced decoherence align favorably or unfavorably with ET energy matching conditions. Parameters: $\lambda_0=50\,\text{cm}^{-1}$, $\hbar\omega_\text{c}=100\,\text{cm}^{-1}$.} 
	\label{fig: collective effect}
\end{figure}

Figure~\ref{fig: collective effect} displays the relative ET rate enhancement $k_\text{cav}/k_\text{non-cav}$ for both $N_\text{mol}=1$ and $N_\text{mol}=2$ across three different driving forces $-\Delta G$. 
The results reveal that cavity-modified ET rates depend sensitively on the number of coupled molecules, confirming the presence of collective effects. 
However, the the nature of this collectivity is not universal---it can be either enhancing or suppressing depending on system parameters.
We identify three distinct regimes of collective behavior depending on the driving force: 
(i) Positive collective effect ($-\Delta G=-100\,\text{cm}^{-1}$, center panel): Enhancement increases with molecular number, following $k_\text{non-cav}<k_\text{cav}(N_\text{mol}=1)<k_\text{cav}(N_\text{mol}=2)$. 
Collective coupling amplifies the cavity's beneficial influence on ET, with the two-molecule system exhibiting stronger rate enhancement than the single-molecule case. 
(ii) Negative collective effect ($-\Delta G=100\,\text{cm}^{-1}$, right panel): Enhancement decreases with molecular number, yielding $k_\text{non-cav}<k_\text{cav}(N_\text{mol}=2)<k_\text{cav}(N_\text{mol}=1)$. 
While both systems exhibit rate enhancement relative to the uncoupled baseline, collective coupling counterintuitively reduces the enhancement magnitude.
The single molecule benefits more from cavity coupling than the two-molecule ensemble.
(iii) Transition behavior [$-\Delta G=-150\,\text{cm}^{-1}$, left panel]: The system changes from cavity-induced suppression to enhancement with increasing $N_\text{mol}$, following $k_\text{cav}(N_\text{mol}=1)<k_\text{non-cav}<k_\text{cav}(N_\text{mol}=2)$. 
For a single molecule, cavity coupling inhibits transfer below the uncoupled rate. 
However, adding a second molecule reverses this effect, restoring enhancement. 
This transition demonstrates that collective effects can qualitatively change the role of the cavity from inhibitor to facilitator.   
These diverse behaviors arise from the interplay of two mechanisms associated with collective polariton formation. 
When $N_\text{mol}$ molecules couple identically to a cavity mode, the effective light-matter coupling strength scales as $\sqrt{N_\text{mol}}$, producing a collective Rabi splitting that grows with ensemble size. 
This enhanced splitting modifies ET dynamics through complementary pathways.
First, the larger Rabi splitting reshapes the polaritonic energy landscape. 
The broader separation between upper and lower polariton branches alters how the system navigates between donor and acceptor states, changing the effective activation barriers and energy matching conditions. 
Second, collective polaritons delocalized across multiple molecules experience reduced coupling to local vibrational environments. 
This spatial delocalization diminishes the effective system-environment interaction per molecular site, enhancing the dynamic polaron decoupling effect and protecting coherent dynamics from decoherence. 
Whether these changes produce positive or negative collective effects depends on their alignment with the ET energy matching condition. 
When the driving force $-\Delta G$ positions the system such that collective Rabi splitting and reduced decoherence both improve energy matching and pathway accessibility, positive enhancement results. 
Conversely, when these modifications shift the system away from optimal resonance conditions or disrupt favorable interference patterns, negative collective effects emerge. 
The transition regime occurs at driving forces where single-molecule cavity coupling creates unfavorable conditions, but collective effects restore or improve energy matching.   

While our HEOM calculations for $N_\text{mol}=1$ and 2 provide rigorous benchmarks, definitive assessment of collective effects in realistic molecular ensembles requires simulating systems with $N_\text{mol}\gg1$. 
Unfortunately, numerically exact quantum dynamics using HEOM becomes computationally prohibitive as Hilbert space dimensions grow exponentially with molecular number.
A promising approach to overcome this limitation employs semiclassical methods such as the truncated Wigner approximation (TWA), which accurately reproduces quantum dynamics in the large-$N$ limit when many molecules are strongly coupled to a cavity mode~\cite{Phuc24, Phuc25}.
Future work combining HEOM benchmarks at small $N_\text{mol}$ with TWA simulations at large $N_\text{mol}$ could provide comprehensive understanding across all experimentally relevant regimes. 
The collective effects studied here connect to broader phenomena in polaritonic chemistry and condensed matter physics, including super-radiant reaction enhancement through constructive interference between spatially separated transfer pathways in molecular ensembles~\cite{Phuc21} and Bose enhancement of transport from polariton Bose-Einstein condensation under strong collective coupling~\cite{Phuc22}.
These diverse manifestations underscore the rich physics accessible when multiple quantum emitters coherently share a photonic environment, revealing collective light-matter coupling as a multifaceted control mechanism whose influence depends critically on the interplay between polaritonic energy structure, decoherence dynamics, and molecular reaction energetics.

\subsection{Generalized Model: Electronic-Vibrational-Photonic Coupling}
The minimal model discussed thus far neglects an important physical effect: the dependence of molecular dipole moments on nuclear coordinates. 
In realistic molecular systems, electronic charge distributions, and hence dipole moments, are modulated by vibrational motion. 
When a molecule vibrates, the spatial overlap between donor and acceptor electronic wavefunctions changes, causing the transition dipole moment $\boldsymbol{\mu}_\text{DA}$ to vary with nuclear displacement. 
Similarly, the diagonal dipole moments $\boldsymbol{\mu}_\text{DD}$ and $\boldsymbol{\mu}_\text{AA}$ are also nuclear-coordinate-dependent. 
Incorporating this coordinate dependence into the cavity coupling leads to a qualitatively new interaction mechanism. 
Expanding the dipole moments to linear order in nuclear displacements and projecting onto the donor-acceptor basis yields a three-body interaction Hamiltonian:
\begin{align}
	\hat{H}_\text{el-vib-ph}=&\eta\left(\hat{a}+\hat{a}^\dagger\right)
	\Big[|\text{D}\rangle\langle\text{D}|+|\text{A}\rangle\langle\text{A}|\nonumber\\
	&+|\text{D}\rangle\langle\text{A}|+|\text{A}\rangle\langle\text{D}|\Big]
	\sum_j \lambda_j \left(\hat{b}_j+\hat{b}_j^\dagger\right),
	\label{eq: three-body interaction}
\end{align} 
where $\eta$ quantifies the strength of this electronic-vibrational-photonic coupling.
This term simultaneously involves the cavity field operator $(\hat{a}+\hat{a}^\dagger)$, electronic state projectors, and vibrational bath coordinates $(\hat{b}_j+\hat{b}_j^\dagger)$, creating a fundamentally nonlinear interaction that cannot be factorized into pairwise couplings.

Physically, Eq.~\eqref{eq: three-body interaction} describes a reciprocal modulation mechanism: vibrational motion alters the effective light-matter coupling strength, while the cavity field simultaneously influences the electronic-vibrational interaction. 
For instance, a vibrational displacement along coordinates coupled to the electronic transition can transiently enhance or suppress the transition dipole moment, thereby modulating the strength of cavity-mediated transfer. 
Conversely, photons in the cavity mode can modify the potential energy surfaces experienced by vibrations, effectively renormalizing vibronic coupling constants. 
This mutual entanglement of electronic, vibrational, and photonic degrees of freedom creates a fundamentally nonlinear interaction that enables quantum interference between multiple transfer pathways, leading to rich dynamical behavior absent in models where cavity and vibrational couplings remain independent.
To make the problem tractable while capturing the essential physics, we adopt several approximations. 
First, we truncate the dipole expansion at linear order in nuclear coordinates, which is justified when zero-point and thermal vibrational amplitudes remain small compared to the characteristic length scales over which the electronic wavefunctions vary. 
Second, we assume equal coupling strengths for all four electronic operators in Eq.~\eqref{eq: three-body interaction}. 
While more refined treatments could distinguish these couplings, this simplification reduces the parameter space while preserving the qualitative physics of three-body entanglement. 
Third, we neglect the dipole self-energy contributions associated with this coordinate-dependent term, an approximation valid outside the ultrastrong coupling regime. 
Finally, we assume that the spectral content of the three-body coupling parallels that of the standard vibronic coupling, characterized by the same spectral density $J(\omega)$. 
This implies that the vibrational modes that couple the donor and acceptor state energies to the bath also modulate the dipole moments with proportional strengths.
While this assumption simplifies the analysis, it represents a physically reasonable scenario where the same nuclear coordinates that determine electronic energy gaps also control charge redistribution and hence dipole moments.

\begin{figure*}
	\centering
	\includegraphics[width=0.9\textwidth]{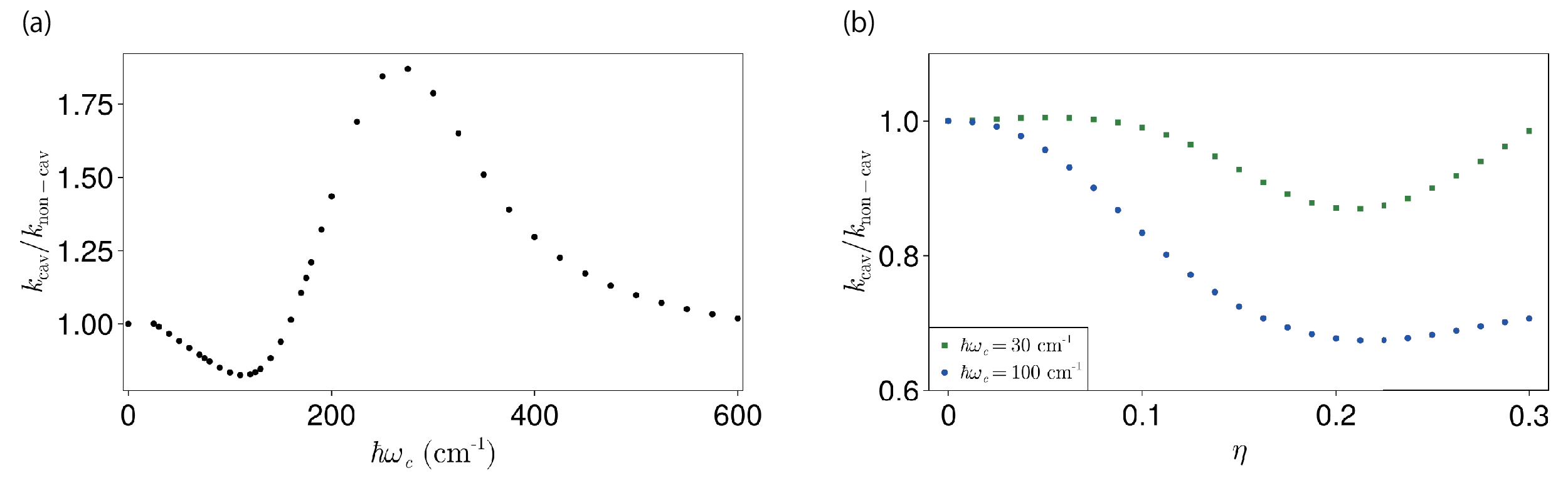}
	\caption{Oscillatory quantum interference in cavity-modified ET with three-body electronic-vibrational-photonic coupling for a Drude-Lorentz environment. ET rate enhancement $k_\text{cav}/k_\text{non-cav}$ as a function of (a) cavity frequency $\omega_\text{c}$ at fixed $\eta=0.1$, and (b) three-body coupling strength $\eta$ at two cavity frequencies: green, $\hbar\omega_\text{c}=30\,\text{cm}^{-1}$; blue, $\hbar\omega_\text{c}=100\,\text{cm}^{-1}$. Standard light-matter couplings are turned off ($t_\text{DA}=g_\text{D,A}=0$) to isolate the three-body effect. The oscillatory, non-monotonic dependencies contrast sharply with the simple resonance and saturation behaviors in the minimal model (cf. Figs.~\ref{fig: dependence on omega_c} and \ref{fig: tDA dependence}), reflecting quantum interference between direct molecular transfer, vibrationally assisted transfer, and cavity-vibration-mediated pathways. Parameters: $\lambda_0=50\,\text{cm}^{-1}$, $-\Delta G=100\,\text{cm}^{-1}$.}
	\label{fig: eta_dependence for Drude-Lorentz}
\end{figure*}

The introduction of three-body electronic-vibrational-photonic coupling fundamentally alters how ET enhancement depends on cavity parameters. 
To isolate the effects of this interaction, we set the standard light-matter couplings to zero ($t_\text{DA}=g_\text{D,A}=0$) so that cavity modification arises solely through the coordinate-dependent term characterized by $\eta$.
Figure~\ref{fig: eta_dependence for Drude-Lorentz} displays the ET rate enhancement as a function of (a) the cavity frequency $\omega_\text{c}$ at fixed $\eta$, and (b) three-body coupling strength $\eta$ for two different cavity frequencies, with the molecular environment modeled as a Drude-Lorentz bath. 
Both dependencies exhibit strikingly different behavior compared to the minimal model. 
Instead of the simple resonance peak observed in the $\omega_\text{c}$ dependence (Fig.~\ref{fig: dependence on omega_c}) or the monotonic-to-saturating trend in the coupling strength dependence (Fig~\ref{fig: tDA dependence}), the three-body interaction induces pronounced oscillatory, non-monotonic behavior in both parameter scans. 
The ET rate alternates between enhancement and suppression as $\omega_\text{c}$ or $\eta$ varies, exhibiting multiple maxima and minima across the explored parameter ranges.
At certain values, the cavity coupling enhances transfer beyond the uncoupled baseline; at others, it suppresses transfer below this baseline despite the presence of light-matter interaction.

This non-monotonic, oscillatory behavior is a clear signature of quantum interference between multiple competing pathways. 
In the presence of the three-body coupling, electron transfer can proceed through several distinct channels: (i) direct electronic tunneling via $H_\text{DA}$; (ii) vibrationally assisted transfer, where bath modes facilitate electronic transitions; and (iii) cavity-vibration-mediated transfer enabled by $\eta$, where photon exchange coupled with nuclear motion creates new transfer routes. 
Each pathway contributes with characteristic amplitude and phase determined by the energy matching condition, coupling strengths, and bath dynamics. 
The overall transfer rate reflects the quantum superposition of these pathways. 
The oscillations arise because varying $\omega_\text{c}$ or $\eta$ tunes the relative phases between interfering pathways. 
When pathways combine constructively, i.e., when their quantum amplitudes add coherently with aligned phases, the transfer rate is exceeds what any single channel could achieve in isolation. 
Conversely, destructive interference occurs when pathway amplitudes partially cancel, reducing the rate below the uncoupled limit. 
The precise conditions for constructive versus destructive interference depend sensitively on multiple factors: the cavity frequency sets the energy scale for virtual photon exchange, the three-body coupling strength $\eta$ determines the amplitude of cavity-vibration-mediated pathways, and the bath spectral density governs how nuclear motion correlates with electronic transitions.
This interference-driven landscape contrasts fundamentally with the minimal cavity-electronic model, where optimization is straightforward: tune the cavity to resonance and increase coupling until saturation. 
With the three-body interaction, optimal ET enhancement requires navigating a complex multidimensional parameter space to identify constructive interference regimes.
This suggests new design principles for cavity-controlled chemistry: rather than simply maximizing coupling strength or matching a resonance condition, one must engineer the relative phases of multiple quantum pathways to achieve desired reactivity outcomes.

To explore how the three-body interaction behaves when a high-frequency intramolecular vibrational mode is involved, we employ a spectral density structure that separates the roles of different vibrational modes. 
Specifically, we assign an underdamped spectral density to the three-body interaction term while retaining the Drude-Lorentz form for the standard vibronic coupling. 
The underdamped spectral density 
\begin{align}
	J_\text{UD}(\omega)=\frac{2\lambda_2^2W_2\omega}{(\omega^2-\omega_0^2)^2+W_2^2\omega^2},
\end{align}
represents a localized vibrational resonance centered at frequency $\omega_0=500\;\text{cm}^{-1}$ with coupling strength $\hbar\lambda_2=\left[1420\,\text{cm}^{-1}\right]^{3/2}$ and damping rate $W_2=\tau_\text{d}^{-1}$, where $\tau_\text{d}=100\,\text{fs}$.
Since $\eta$ serves as an overall amplitude factor that is now absorbed into the spectral density normalization, we set $\eta=1$ in this analysis.
This model captures a physically motivated scenario: high-frequency intramolecular vibrations modulate the molecular transition dipole moment and thus couple to the cavity field through the three-body term, while low-frequency intermolecular vibrations described by the Drude-Lorentz bath primarily couple to the electronic state energies through standard vibronic interactions.
The narrow linewidth of the underdamped mode (set by $W_2$) creates a well-defined spectral feature that, in simple resonance models, would be expected to produce a pronounced peak when matched with the cavity frequency.

\begin{figure}
	\centering
	\includegraphics[width=0.9\linewidth]{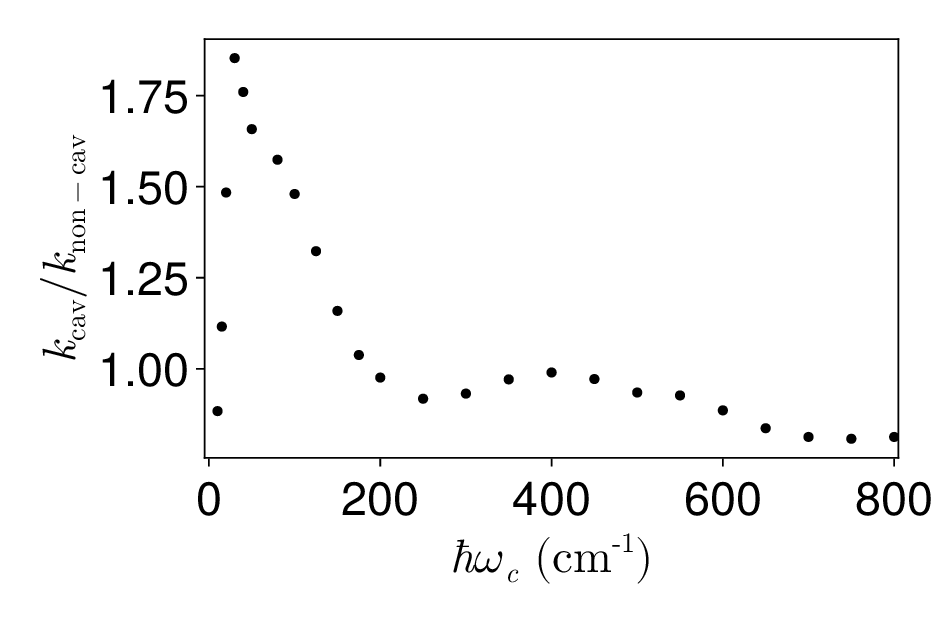}
	\caption{ET rate enhancement $k_\text{cav}/k_\text{non-cav}$ as a function of cavity frequency $\omega_\text{c}$ for a system where a high-frequency underdamped vibrational mode ($\omega_0=500\;\text{cm}^{-1}$, $\hbar\lambda_2=\left[1420\,\text{cm}^{-1}\right]^{3/2}$, $\tau_\text{d}=100\,\text{fs}$) couples through the three-body interaction to the cavity field, while a Drude-Lorentz bath ($\lambda_0=50\,\text{cm}^{-1}$, $\tau_\text{b}=100\,\text{fs}$) couples to the molecular electronic states. The absence of a simple resonance peak at $\hbar\omega_\text{c}\simeq\hbar\omega_0$ and the presence of a complex oscillatory structure---with the primary maximum shifted to $\hbar\omega_\text{c}\simeq 30\,\text{cm}^{-1}$---demonstrate that three-body electronic-vibrational-photonic coupling induces quantum interference between multiple transfer pathways rather than conventional vibrational polariton resoannce. Other parameters: $-\Delta G=100\,\text{cm}^{-1}$.}
	\label{fig: underdamped added to Drude Lorentz}
\end{figure}

Figure~\ref{fig: underdamped added to Drude Lorentz} displays the ET rate enhancement $k_\text{cav}/k_\text{non-cav}$ as a function of cavity frequency.
In a conventional cavity-vibration coupling scenario, one would anticipate straightforward resonance behavior: when the cavity frequency approaches the vibrational mode frequency ($\hbar\omega_\text{c}\simeq\hbar\omega_0=500\,\text{cm}^{-1}$), strong vibrational polariton formation should occur, enabling efficient resonant energy exchange between cavity photons and nuclear motion.
This resonance would manifest as a prominent peak in the enhancement centered near $\hbar\omega_\text{c}\simeq\hbar\omega_0$ with a width determined by the vibrational damping. 
However, the observed frequency dependence deviates dramatically from this simple expectation.
Instead of a single resonance peak at $\hbar\omega_\text{c}=500\,\text{cm}^{-1}$, the enhancement exhibits a complex multi-featured profile. 
The primary maximum appears at $\hbar\omega_\text{c}\simeq 30\,\text{cm}^{-1}$, dramatically shifted downward by approximately $\simeq 470\,\text{cm}^{-1}$ from the bare vibrational mode frequency. 
This large peak is followed by an oscillatory tail extending to higher frequencies, exhibiting alternating regions of enhancement and suppression.
Remarkably, certain frequency ranges show $k_\text{cav}/k_\text{non-cav}<1$, indicating that the cavity coupling actually suppresses ET below the uncoupled baseline despite nominally providing an additional pathway for energy redistribution.

This nonresonant, oscillatory structure---similar to the behavior observed with the Drude-Lorentz spectral density for the three-body interaction (Fig.~\ref{fig: eta_dependence for Drude-Lorentz}a)---provides compelling evidence that quantum interference, rather than classical resonant energy matching, governs the system's dynamics.
The absence of a pronounced resonance at $\hbar\omega_\text{c}\simeq\hbar\omega_0$ is particularly revealing. 
It demonstrates that the vibrational mode's contribution cannot be understood in isolation as a simple single mode resonantly exchanging energy quanta with the cavity. 
Rather, the mode's effect is thoroughly entangled with all other system dynamics through the nonlinear three-body term. 
The electronic states, continuous Drude-Lorentz bath modes, discrete high-frequency vibrational mode, and cavity photons form a deeply correlated quantum many-body system where conventional distinctions between resonant and off-resonant contributions lose meaning. 
Under this vibronic polariton formation regime, ET dynamics emerge from collective interference among all available pathways in this coupled electronic-vibrational-photonic landscape. 
Accurate description demands treating all degrees of freedom on equal footing rather than attempting perturbative inclusion of individual resonances within a separable framework.

\section{Conclusion}
We have investigated electron transfer dynamics in cavity-coupled molecular systems using numerically exact HEOM simulations, revealing both fundamental insights and control mechanisms for polaritonic chemistry.
Beginning with the minimal model, in which cavity fields couple exclusively to electronic degrees of freedom, we examined two distinct light-matter coupling channels: direct transition coupling ($t_\text{DA}$) and energy-fluctuation coupling ($g_\text{D,A}$).
Both channels enhance ET rates across wide ranges of driving forces and reorganization energies, with direct transition coupling generally providing stronger enhancement by creating coherent transfer pathways through virtual photon exchange. 
This acceleration is accompanied by extended coherence times, consistent with the dynamic polaron decoupling effect whereby strong light-matter coupling suppresses vibronic decoherence. 
The ET rate exhibits resonance behavior in both channels, though optimal cavity frequencies differ due to their distinct physical mechanisms: the $t_\text{DA}$ channel resonates when the cavity matches the effective electronic transition energy, while the $g_\text{D,A}$ channel's resonance reflects optimal modulation of the activation barrier.

By comparing HEOM results against perturbative predictions from Fermi's golden rule (FGR), we found that while FGR captures qualitative trends, it fails quantitatively in the strong coupling regime: the predicted quadratic scaling of the rate with coupling strength breaks down beyond a threshold, and even in the moderate coupling regime, where quadratic scaling approximately holds, the prefactor differs significantly from the perturbative prediction. 
This saturation behavior underscores the necessity of nonperturbative methods for accurate quantitative predictions in the strong coupling limit. 
We identified additional phenomena within the minimal model framework. 
Collective effects emerge when multiple molecules couple to a single cavity mode: the ET rate changes with molecular number even at fixed per-molecule coupling strength, exhibiting parameter-dependent enhancement or suppression depending on how collective Rabi splitting and delocalization-induced decoherence protection align with energy matching conditions. 
Cavity loss introduces competing mechanisms: moderate dissipation enhances transfer through fluctuation-assisted processes and resonance broadening, while excessive loss destroys the coherent polariton dynamics underlying cavity-mediated enhancement, resulting in non-monotonic rate dependence on the photon decay rate.

Our generalized model, incorporating the nuclear-coordinate dependence of transition dipole moments, unveils qualitatively new physics. 
This extension introduces a three-body electronic-vibrational-photonic interaction that fundamentally alters the nature of cavity control. 
Rather than the monotonic or saturating behaviors of the minimal model, the three-body coupling induces oscillatory, non-monotonic dependence of ET rates on both cavity frequency and coupling strength. 
These oscillations arise from quantum interference between multiple transfer pathways, whose relative phases vary with system parameters.
We demonstrated this interference-driven dynamics for both Drude-Lorentz environments representing continuous low-frequency modes and composite spectral densities including discrete high-frequency vibrations. 
Remarkably, even when a well-defined intramolecular mode couples through the three-body term, the ET enhancement exhibits complex multi-featured frequency dependence with the primary maximum shifted far from the vibrational resonance and followed by an oscillatory tail. 
This absence of simple resonance behavior confirms that nonlinear quantum interference, rather than conventional energy matching, governs the dynamics. 

These findings establish that optical cavities provide a dual control mechanism for molecular reactivity: tuning electronic energy gaps via excitonic polaritons and modulating reaction pathways through vibronic polaritons. 
The oscillatory interference effects we identify suggest new strategies for cavity-mediated chemical control that exploit quantum coherence and many-body correlations, opening avenues for designing cavity-enhanced molecular devices and catalysts.


\begin{acknowledgements}
The authors thank Hirofumi Sato for fruitful discussions.	

N. T. Phuc acknowledges financial support from JST-PRESTO program: Exploring Quantum Frontiers Through Quantum-Classical Interdisciplinary Fusion (JPMJPR24F2) and Hirose Foundation.
T. Fukushima acknowledges financial support from JST-PRESTO program: Exploring Quantum Frontiers Through Quantum-Classical Interdisciplinary Fusion (JPMJPR23FB).

The computational work was conducted using the facilities of the Research Center for Computational Science, Okazaki, Japan.
\end{acknowledgements}

\section*{Author Declarations}
\subsection*{Conflict of Interest}
The authors have no conflicts to disclose.

\subsection*{Author Contributions}
Takumi Hidaka: Formal analysis (supporting); Investigation (equal); Methodology (supporting); Writing – original draft (supporting); Writing – review \& editing (supporting). 
Tomohiro Fukushima: Formal analysis (supporting); Funding acquisition (supporting); Writing – review \& editing (supporting).
Nguyen Thanh Phuc: Conceptualization; Formal analysis (lead); Investigation (equal); Funding acquisition (lead); Methodology (lead); Writing – original draft (lead); Writing – review \& editing (lead). 

\subsection*{Data availability}
The data that support the findings of this study are available from the corresponding author upon reasonable request.




\end{document}